\documentclass[apj]{emulateapj}
\usepackage{natbib}
\usepackage{apjfonts}
\usepackage{mathptmx}

\newcommand  \kms      {\ifmmode {\rm km\,s}^{-1} \else km\,s$^{-1}$\fi}

\newcommand  \cmii     {\hbox{cm$^{-2}$}}
\newcommand  \ergs     {\ifmmode {\rm ergs\,s}^{-1} \else ergs s$^{-1}$\fi}
\newcommand  \ergcms   {\ifmmode {\rm ergs\,cm}^{-2}\,{\rm s}^{-1}
                        \else ergs\,cm$^{-2}$\,s$^{-1}$\fi}
\newcommand  \ergcmsA {\ifmmode{\rm ergs\,cm}^{-2}\,{\rm s}^{-1}\,{\rm\AA}^{-1}
                        \else ergs\,cm$^{-2}$\,s$^{-1}$\,\AA$^{-1}$\fi}
\newcommand \ergcmsHz {\ifmmode{\rm ergs\,cm}^{-2}\,{\rm s}^{-1}\,{\rm Hz}^{-1}
                        \else ergs\,cm$^{-2}$\,s$^{-1}$\,Hz$^{-1}$\fi}
\newcommand  \phcms    {\ifmmode {\rm ph\,cm}^{-2}\,{\rm s}^{-1}
                        \else ,ph\,cm$^{-2}$\,s$^{-1}$\fi}
\newcommand  \phcmsA   {\ifmmode {\rm ph\,cm}^{-2}\,{\rm s}^{-1}\,{\rm\AA}^{-1}
                        \else ph\,cm$^{-2}$\,s$^{-1}$\,\AA$^{-1}$\fi}
\def\micron{\ifmmode \mu{\rm m} \else $\mu$m\fi}
\def\kms{\ifmmode {\rm km\,s}^{-1} \else km\,s$^{-1}$\fi}
\def\Hubble{\ifmmode {\rm km\,s}^{-1}\,{\rm Mpc}^{-1}
        \else km\,s$^{-1}$\,Mpc$^{-1}$\fi}
\def\ergsec{\ifmmode {\rm ergs\;s}^{-1} \else ergs s$^{-1}$\fi}
\def\ergscm{\ifmmode {\rm ergs\,s}^{-1}\,{\rm cm}^{-2}
          \else ergs\,s$^{-1}$\,cm$^{-2}$\fi}
\def\ergscmA{\ifmmode {\rm ergs\,s}^{-1}\,{\rm cm}^{-2}\,{\rm \AA}^{-1}
          \else ergs\,s$^{-1}$\,cm$^{-2}$\,\AA$^{-1}$\fi}
\def\ergscmHz{\ifmmode {\rm ergs\,s}^{-1}\,{\rm cm}^{-2}\,{\rm Hz}^{-1}
          \else ergs\,s$^{-1}$\,cm$^{-2}$\,Hz$^{-1}$\fi}
\def\Msun{\ifmmode M_{\odot} \else $M_{\odot}$\fi}
\def\Lsun{\ifmmode L_{\odot} \else $L_{\odot}$\fi}
\def\qo{\ifmmode q_{0} \else $q_{0}$\fi}
\def\Ho{\ifmmode H_{0} \else $H_{0}$\fi}
\def\ho{\ifmmode h_{0} \else $h_{0}$\fi}
\def\qo{\ifmmode q_{0} \else $q_{0}$\fi}
\def\ao{\ifmmode a_{0} \else $a_{0}$\fi}
\def\to{\ifmmode t_{0} \else $t_{0}$\fi}
\def\Halpha{\ifmmode {\rm H}\alpha \else H$\alpha$\fi}
\def\Hbeta{\ifmmode {\rm H}\beta \else H$\beta$\fi}
\def\hb{\ifmmode {\rm H}\beta \else H$\beta$\fi}
\def\Hgamma{\ifmmode {\rm H}\gamma \else H$\gamma$\fi}
\def\Hdelta{\ifmmode {\rm H}\delta \else H$\delta$\fi}
\def\Lya{\ifmmode {\rm Ly}\alpha \else Ly$\alpha$\fi}
\def\Lyb{\ifmmode {\rm Ly}\beta \else Ly$\beta$\fi}
\def\hi{\ifmmode \mbox{{\rm H}\,{\sc i}} \else H\,{\sc i}\fi}

\def\ciii{\ifmmode {\rm C}\,{\sc iii} \else C\,{\sc iii}\fi}

\def\oiii{[O\,{\sc iii}]\,$\lambda5007$}

\def\o5007{[O\,{\sc iii}]\,$\lambda5007$}
\def  \RNLR        {\hbox{$ {R_{\rm NLR}} $}}

\def  \kms         {\hbox{km s$^{-1}$}}          
\def  \ergs        {\hbox{erg s$^{-1}$}}              

\def  \cmii        {\hbox{cm$^{-2}$}}

\def  \La          {\ifmmode {\rm Ly}\alpha \else Ly$\alpha$\fi}
\def  \Ka          {\ifmmode {\rm K}\alpha \else K$\alpha$\fi}
\def  \Lb          {\ifmmode {\rm L}\beta \else L$\beta$\fi}
\def  \Ha          {\ifmmode {\rm H}\alpha \else H$\alpha$\fi}
\def  \Hb          {\ifmmode {\rm H}\beta \else H$\beta$\fi}
\def  \Pa          {\ifmmode {\rm P}\alpha \else P$\alpha$\fi}
\def  \CIIIb       {\ifmmode {\rm C}\,{\sc iii]}\,\lambda1909
                     \else C\,{\sc iii]}\,$\lambda1909$\fi}
\def  \CIV         {\ifmmode {\rm C}\,{\sc iv}\,\lambda1549
                     \else C\,{\sc iv}\,$\lambda1549$\fi}
\def  \MgII         {\ifmmode {\rm Mg}\,{\sc ii}\,\lambda2798
                     \else Mg\,{\sc ii}\,$\lambda2798$\fi}
\def  \OVI         {\ifmmode {\rm O}\,{\sc vi}\,\lambda1035
x
                     \else O\,{\sc vi}\,$\lambda1035$\fi}
\shortauthors{NETZER ET AL.}
\shorttitle{DISAPPEARING NARROW LINE REGIONS AND THE ROLE OF ACCRETION}

\journalinfo{The Astrophysical Journal, 614:???--???, 2004 October 10,
astro-ph/0406560}

\slugcomment{Received 2004 March 28; accepted 2004 June 22}

\begin{document}

\title{Near Infrared Spectroscopy of High Redshift Active Galactic Nuclei. \\
II. Disappearing Narrow Line Regions and the Role of Accretion}

\author{
H.~Netzer,\altaffilmark{1}
O.~Shemmer,\altaffilmark{1}
R.~Maiolino,\altaffilmark{2}
E.~Oliva,\altaffilmark{3}
S.~Croom,\altaffilmark{4}
E.~Corbett,\altaffilmark{4}
and L.~di Fabrizio \altaffilmark{3}
}

\altaffiltext{1}
                {School of Physics and Astronomy and the Wise
                Observatory, The Raymond and Beverly Sackler Faculty of
                Exact Sciences, Tel-Aviv University, Tel-Aviv 69978,
                Israel; netzer@wise.tau.ac.il}

\altaffiltext{2}
                {INAF - Osservatorio Astrofisico di Arcetri, Largo E. Fermi 5,
                I-50125 Firenze, Italy; maiolino@arcetri.astro.it}

\altaffiltext{3}
                {Istituto Nazionale di Astrofisica, Centro Galileo Galilei, and
                Telescopio Nazionale Galileo, P.O. Box 565, E-38700
                Santa Cruz de la Palma, Spain; oliva@tng.iac.es}

\altaffiltext{4}
                {Anglo-Australian Observatory, PO Box 296, Epping, NSW
                1710, Australia; scroom@aaoepp.aao.gov.au}


\begin{abstract}
We present new near infrared spectroscopic measurements for 29 luminous
high-redshift active galactic nuclei (AGNs) and use the data to discuss the
size and other properties of the narrow line regions (NLRs) in those sources. 
The high resolution spectra have been used to carefully model the
\ion{Fe}{2} blends and to provide reliable \oiii, Fe {\sc ii} and \hb\
measurements.
We find that about 2/3 of all very high luminosity sources show 
strong \oiii\ lines while the remaining objects show no or very weak such line.
While weak \oiii\ emitters are also found among lower luminosity AGNs, we
argue that the implications for very high luminosity objects are different.
In particular, we suggest that the averaging of these two populations in other
works gave rise to claims of a Baldwin relationship in \oiii\ which is not
confirmed by our data. We also argue that earlier proposed relations of the
type \RNLR$\propto L_{{\rm [O \ III]}}^{1/2}$, where \RNLR\ is the radius
of the NLR, are theoretically sound yet they must break down for \RNLR\
exceeding a few kpc. This suggests that the NLR properties in high luminosity
sources are very different from those observed in nearby AGNs. In particular,
we suggest that some sources lost their very large, dynamically unbound NLR
while others are in a phase of violent star-forming events that produce a large
quantity of high density gas in the central kpc. This gas is ionized
and excited by the central radiation source and its spectroscopic properties
may be different from those observed in nearby, lower luminosity NLRs.
We also discuss the dependence of EW(\hb) and \ion{Fe}{2}/\hb\ on luminosity,
black hole mass, and accretion rate for a large sample of AGNs. The strongest
dependence of the two quantities is on the accretion rate and the
\ion{Fe}{2}/\hb\ correlation is probably due to the EW(\hb) dependence on
accretion rate. We show the most extreme values measured so far of
\ion{Fe}{2}/\hb\ and address its correlation with EW(\oiii).
\end{abstract}


\section{Introduction}
\label{introduction}

The narrow line regions (NLRs) of active galactic nuclei (AGNs) have been
studied, extensively, from the ground and from space. This component
of the nucleus is spatially resolved from the ground in nearby sources
and {\sl HST} observations extend the range to a redshift of about 0.5.
Thus, detailed NLR mappings are now available for a large number of
sources covering a large range of luminosity and redshift
(Falcke, Wilson, \& Simpson 1998; Bennert et al. 2002, hereafter B02,
Schmitt et al. 2003).

The spectroscopic characteristics of the NLR gas have been studied,
extensively, over several decades and high quality data are now available 
(e.g., Veilleux \& Osterbrock 1987).
The main source of excitation of the NLR gas is photoionization by the
the central continuum (see review and references in Netzer 1990) but
shock excitation must be important in some parts of this region, most
notably is NLRs that are associated with jet-like radio structures
(Schiano 1986; Dopita et al. 2002 and references therein). The gas
dynamics has been studied too with detailed results
concerning the profiles of various emission lines and their dependence on 
the level of ionization, the density and the dust content of the gas  
(e.g., Veilleux 1991; Nelson \& Whittle 1996; Barth et al., 2001).
A major emphasis in recent years has been the extension
of such works to higher luminosity sources.
Some such studies suggest that the \oiii\ line width is
correlated with the stellar velocity distribution in the bulge and
thus also with the mass of the central black hole (Nelson, 2000;
Shields et al., 2003).

Several recent attempts to study NLRs in large samples of AGNs lead to
apparently conflicting results. B02 obtained narrow band HST images of
seven luminous radio-quiet Palomar-Green (PG) quasars with $z<0.5$. They
argued, on the basis of comparison with nearby less luminous sources, that the
NLR size (radius) scales with the \oiii\ and the \hb\ line luminosities
roughly as \RNLR$\propto L^{0.5}$. The measured NLR sizes in their
most luminous sources approached 10 kpc (throughout this work
we assume ${\rm H}_0=70$ \kms\ Mpc$^{-1}$, $\Omega_{\rm m}=0.3$, and
$\Omega_{\Lambda}=0.7$).
This dependence has been questioned by Schmitt et al. (2003) who
studied a much larger sample (22 Seyfert 1's and 38 Seyfert 2's),
albeit with much lower luminosity, and found \RNLR$\propto L^{0.33}$.
Croom et al. (2002) analyzed the spectra of $\sim 22,000$ AGNs from
the 2dF quasar redshift survey (2QZ) and claimed to see a decrease in
the equivalent width (EW) of several narrow lines ([\ion{O}{2}]$\lambda3727$,
[\ion{Ne}{5}]$\lambda3426$, and [\ion{Ne}{3}]$\lambda3870$) with source
luminosity. 
They suggested that at least part of this ``Baldwin effect'' (Baldwin
1977) is due to the increase in NLR size with source luminosity which
leads to galactic-scale dimensions in the most luminous objects. Such
NLRs are likely to escape the system leading to AGNs with weak or no NLR
emission. Testing this idea for the most intense narrow line, \oiii,
was limited by the low redshift, and hence relatively low luminosity in the
Croom et al. (2002) ground-based sample. 

This paper addresses the issues of ``the disappearing NLRs'' and the
\ion{Fe}{2}/\hb\ ratio in high
luminosity AGNs. The work complements the Shemmer et al. (2004;
hereafter Paper I) study and is based on the same data set. The paper
is arranged as follows: \S~\ref{observations} presents the new observations,
\S~\ref{LEWR} shows various correlations involving the \oiii, \hb, and Fe
{\sc ii} lines and \S~\ref{discussion} discusses the implications regarding the
size and the physics of the NLR as well as the \ion{Fe}{2}/\hb\ ratio in the
most luminous AGNs. 

\begin{deluxetable*}{lccccccccccc}
\tablecolumns{12}
\tablewidth{0pt}
\tablecaption{Continuum and Emission Line Measurements. \label{fit}}

\tablehead
{
\colhead{Quasar Name} &
\colhead{z\tablenotemark{a}} &
\colhead{Log $\lambda L_{\lambda}(5100)$} &
\multicolumn{2}{c}{EW([O {\sc iii}])}&
\multicolumn{2}{c}{$L_{[\rm O \ III]}$} &
\multicolumn{2}{c}{FWHM([O {\sc iii})} &
\colhead{$L_{{\rm H}\beta}$} &
\colhead{Fe {\sc ii}/H$\beta$} \\
\colhead{} &
\colhead{} &
\colhead{} &
\colhead{Best Fit} &
\colhead{Direct} &
\colhead{Best Fit} &
\colhead{Direct} &
\colhead{Best Fit} &
\colhead{Direct} &
\colhead{Best Fit} &
\colhead{} \\
\colhead{} &
\colhead{} &
\colhead{[erg s$^{-1}$]} &
\colhead{[\AA]} &
\colhead{[\AA]} &
\colhead{[erg s$^{-1}$]} &
\colhead{[erg s$^{-1}$]} &
\colhead{[\kms]} &
\colhead{[\kms]} &
\colhead{[erg s$^{-1}$]} &
\colhead{} \\
\colhead{(1)} &
\colhead{(2)} &
\colhead{(3)} &
\colhead{(4)} &
\colhead{(5)} &
\colhead{(6)} &
\colhead{(7)} &
\colhead{(8)} &
\colhead{(9)} &
\colhead{(10)} &
\colhead{(11)}
}
\startdata
2QZ J001221.1$-$283630      &2.339& 46.26 &$<10$&\nodata &$<43.59$&\nodata
&\nodata &\nodata & 44.29 & 1.27 \\
2QZ J002830.4$-$281706      &2.401& 46.58 & 32  & 31     & 44.40  & 44.38  &
1511   & 1823   & 44.63 & 0.37 \\
UM 667                      &3.132& 46.28 & 16  & 16     & 43.78  & 43.79  &
759    & 1468   & 44.44 & 2.08 \\
LBQS 0109$+$0213            &2.349& 46.80 & 25  & 31     & 44.50  & 44.59  &
1398   & 1341   & 44.93 & 0.14 \\
\protect{[HB89]} 0123$+$257 &2.369& 46.57 & 27  & 27     & 44.31  & 44.32  &
587   &  532   & 44.75 & 0.26 \\
2QZ J023805.8$-$274337      &2.471& 46.57 & $<7$&\nodata &$<43.70$&\nodata &
\nodata &\nodata & 44.70 & 1.57 \\
SDSS J024933.42$-$083454.4  &2.491& 46.38 & 27  & 27     & 44.11  & 44.10  &
815   &  394   & 44.54 &$<0.1$\\
\protect{[HB89]} 0329$-$385 &2.435& 46.71 & 20  & 24     & 44.30  & 44.37  &
491   &  450   & 44.82 & 0.30 \\
\protect{[HB89]} 0504$+$030 &2.473& 46.32 & 73  & 74     & 44.49  & 44.51  &
1065   &  836   & 44.38 & 0.49 \\
SDSS J100428.43$+$001825.6  &3.046& 46.44 & 54  & 60     & 44.46  & 44.51  &
527    &  996   & 44.76 & 0.59 \\
TON 618                     &2.226& 47.31 & $<3$&\nodata &$<44.11$&\nodata &
\nodata &\nodata & 45.42 & 0.65 \\
\protect{[HB89]} 1246$-$057 &2.240& 47.16 & $<5$&\nodata &$<44.14$&\nodata &
\nodata &\nodata & 45.14 & 1.20 \\
\protect{[HB89]} 1318$-$113 &2.306& 46.89 & 14  & 13     & 44.31  & 44.28  &
1903   & 2303   & 44.84 &$<0.1$\\
\protect{[HB89]} 1346$-$036 &2.370& 46.88 & $<3$&\nodata &$<43.72$&\nodata &
\nodata &\nodata & 45.02 & 0.87 \\
SDSS J135445.66$+$002050.2  &2.531& 46.49 & $<3$&\nodata &$<43.22$&\nodata &
\nodata &\nodata & 44.52 & 0.63 \\
UM 629                      &2.460& 46.56 & 35  & 23     & 44.40  & 44.23  &
1413   &  929   & 44.77 & 1.29 \\
UM 632                      &2.517& 46.54 & 15  & 16     & 44.03  & 44.05  &
196   &  492   & 44.85 & 0.35 \\
UM 642                      &2.361& 46.29 & 11  & 14     & 43.63  & 43.74  &
1696   & 1859   & 44.53 & 0.41 \\
UM 645                      &2.257& 46.31 & 23  & 30     & 43.98  & 44.09  &
525   &  633   & 44.67 & 0.10 \\
SBS 1425$+$606              &3.202& 47.38 & 23  & 19     & 45.03  & 44.96  &
1382   & 659    & 45.37 & 0.34 \\
SDSS J170102.18$+$612301.0  &2.301& 46.34 & $<7$&\nodata &$<43.50$&\nodata &
\nodata &\nodata & 44.50 & 1.06 \\
SDSS J173352.22$+$540030.5  &3.428& 47.00 & 11  & 8      & 44.34  & 44.22  &
1335   & 833    & 44.99 & 0.39 \\
\protect{[HB89]} 2126$-$158 &3.282& 47.25 & 13  & 9      & 44.64  & 44.51  &
1327   & 1040   & 45.47 & 0.49 \\
\protect{[HB89]} 2132$+$014 &3.199& 45.77 & 59  & 52     & 43.83  & 43.77  &
1315   & 1328   & 44.27 & 0.54 \\
2QZ J221814.4$-$300306      &2.389& 46.54 & 12  & 13     & 43.94  & 43.96  &
1791   & 2343   & 44.54 & 0.57 \\
2QZ J222006.7$-$280324      &2.414& 47.22 & 13  & 16     & 44.63  & 44.73  &
1019   &  908   & 45.14 & 0.42 \\
\protect{[HB89]} 2254$+$024 &2.083& 46.45 & 15  & 13     & 43.94  & 43.86  &
612   & 1390   & 44.78 & 1.27 \\
2QZ J231456.8$-$280102      &2.400& 46.31 & 15  & 15     & 43.79  & 43.81  &
1267   & 1654   & 44.47 &$<0.1$\\
2QZ J234510.3$-$293155      &2.382& 46.32 & 20  & 28     & 43.96  & 44.11  &
887   &  758   & 44.79 & 0.82 \\
\enddata
\tablecomments{The methods used to obtain 'Best Fit' and 'Direct'
measurements are outlined in Paper I.}
\tablenotetext{a}{Systemic redshift (see Paper I).}
\end{deluxetable*}

\section{Observations}
\label{observations}

We obtained new near infrared (IR) spectroscopic observations for a sample of
29 high-redshift, high-luminosity AGNs at the Anglo-Australian Telescope (AAT)
in Australia and at Telescopio Nazionale Galileo (TNG) in Spain.
The observations and data reduction are described in Paper I
where all IR spectra are also shown (their Figures 1--3).

In this work we focus on the \oiii\ emission line that, when observed,
is the strongest narrow line in the spectrum (all numbers given
in this paper refer to the 5007\AA\ component {\it only}).
We also show and discuss various correlations involving the broad
\ion{Fe}{2} lines.
The measurement of the \oiii\ line is complicated due to the presence of
strong, broad \ion{Fe}{2} blends in this part of the spectrum. The accurate
modeling of these blends is crucial to our study of the \ion{Fe}{2}
spectrum as well as the measurement of the \oiii\ line. Below we give a
detailed description of this process and illustrate the results for
the case of [HB89]~1346$-$036.

\begin{figure}
\epsscale{1.2}
\plotone{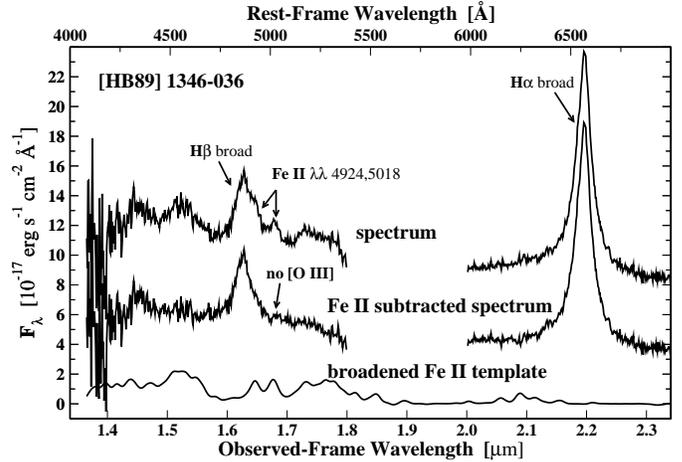}
\caption
{An example of the \oiii\ line measurement process for the z=2.370
quasar \protect{[HB89]} 1346$-$036 we observed at TNG in 2002. The
diagram shows the reduced calibrated spectrum with emission lines
resembling the \oiii\ doublet ({\it top curve}) and the \ion{Fe}{2}
subtracted spectrum ({\it middle curve}) where no sign of \oiii\ is seen.
The \ion{Fe}{2} template is also shown ({\it bottom curve}).}
\label{q1346}
\end{figure}

The near-IR spectrum of the quasar [HB89]~1346$-$036 (z=2.370) was observed by
McIntosh et al. (1999). The same observation was later used by Yuan \& Wills
(2003) who remeasured the McIntosh et al. (1999) spectrum and used it in their
study of the Eddington ratio in $z\sim2$ quasars. Both McIntosh et al. (1999)
and Yuan \& Wills (2003) found an \oiii\ line in this source, which
was later used in several of their correlations. Our superior, high
S/N, better resolution IR spectroscopy of the source shows broad lines
due to \hb\ and \ion{Fe}{2} and a weak emission feature very close to
the putative \oiii\ position (Fig.~\ref{q1346} top curve). We have
used the empirical \ion{Fe}{2} emission template of Boroson \& Green (1992;
hereafter BG92), scaled to the intensity of the strongest iron features in
order to remove those \ion{Fe}{2} lines. The broadened template is shown in
the diagram and the \ion{Fe}{2} subtracted spectrum is given below the original
spectrum. As seen in the diagram, the process completely removed
all trace of \oiii\ emission. We applied a similar procedure to the
spectra of all other sources and the results listed below are all
corrected for the \ion{Fe}{2} blends. We suspect that other sources in
the McIntosh et al. (1999) and the Yuan and Wills (2003) samples
suffer from a similar problem and hence decided not to include these objects
in our analysis. We note that Sulentic et al. (2004) also
question some of the McIntosh et al. (1999) measurements.

Table~\ref{fit} gives a summary of the data used in this paper. It includes
the object
name (column 1), systemic redshift (column 2), continuum luminosity defined as
$\lambda L_{\lambda}$ at rest wavelength 5100\AA\ (column 3) and the
basic \oiii\ line measurements (rest-frame EW in columns 4 \& 5,
luminosity in columns 6 \& 7, and FWHM in columns 8 \& 9). Table~\ref{fit}
also gives the best-fit \hb\ luminosity in column 10, and the Fe {\sc ii}/\hb\
flux ratio\footnote{This ratio matches the BG92 definition of R Fe {\sc ii},
i.e., the ratio between the EW of the Fe {\sc ii} blends in the
$\lambda4434$--$\lambda4684$ band and EW(\hb)} in column 11.
Regarding the uncertainties on those numbers,
some of those are discussed in Paper I and the others, related to the
\oiii\ line, were obtained using the procedure explained in Paper I as
applied to this line.

The main result, which is apparent in Table~\ref{fit}, is that the
population of high-redshift, high-luminosity quasars is divided into
two distinct groups. One group (22 sources) contains objects with strong
\oiii\ lines (EW$\sim 10-80$\,\AA) and with  I(\oiii)/I(broad \hb)
similar to the ratio observed in many low luminosity type-I AGNs.
The second group (seven sources) shows {\it no} \oiii\ {\it line} within the
observational uncertainty. To obtain the upper limits on 
EW(\oiii) in those sources, we assumed a ``typical'' \oiii\ line
with FWHM of 1000 \kms\ (about the median in our sample, see below) and
looked for the weakest such feature that would have been detected in
our spectra after the removal of the \ion{Fe}{2} blends. For the best
S/N spectra (three sources), this translates to an EW which is approximately
$0.05 \times$EW(\hb). For the other four cases the upper limits
correspond to $0.1-0.2 \times$EW(\hb). The luminosities corresponding to 
these upper limits are listed in Table~\ref{fit}.
 
The division into two distinct groups of very high luminosity AGNs
is further confirmed by the Dietrich et al. (2002a) observations. These authors
found that two out of the six luminous $z\simeq3.5$ quasars in their sample
have prominent \oiii\ lines while the remaining four had no trace of this
line. Combining with our new data we find that out of 35 high luminosity
quasars, 24 show strong \oiii\ and 11 others are consistent with no such line
in their spectrum.

\begin{figure}
\epsscale{1.2}
\plotone{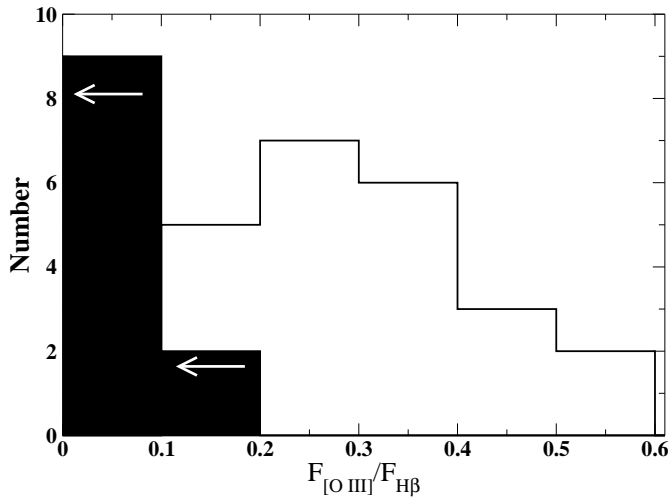}
\caption
{\oiii/\hb\ line ratio histogram for our sources.
The dark region indicates sources with upper limits ({\it arrows}) on
the \oiii\ flux.}
\label{O3hb_hist}
\end{figure}

We have checked this finding in various ways. In particular, we have
examined the distribution of I(\oiii)/I(\hb) which is shown
in Fig.~\ref{O3hb_hist}. The histogram is made up of a broad distribution
centered at about 0.3 and a group of sources with I(\oiii)/I(\hb)$<0.1$.
Unfortunately, the uncertainty on the upper limit of I(\oiii) in two
of the sources is very large and those objects bridge the gap between the
strong and the very weak \oiii\ emitters. We must therefore consider two
hypotheses: one is a real dichotomy in the \oiii\ line intensity and
the other, a continuous distribution in EW(\oiii) which has a long
tail at very small EWs.

Very high luminosity sources that are also weak \oiii\ emitters have been
found by Yuan and Wills (2003) in their IR study of high redshift quasars.
However, most of those objects are broad absorption line quasars (BALQSOs)
that are known to have a weak \oiii\ line. Regarding intermediate luminosity
AGNs, BG92 find that 26 out of the 87 sources in their PG
quasar sample show EW(\oiii)$<10$\AA\ and 12 show EW(\oiii)$<6$\AA.
In addition, about half of the quasars in the new Sulentic et al. (2004)
high-$z$ sample have EW(\oiii)$<6$\AA.
Assuming the upper limits we obtained can be translated to actual EW
measurements, we conclude that the EW distribution in our sample
is not very different from that of BG92.
Combining with the information about nearby Seyfert 1s we conclude that
there are very few weak \oiii\ emitters among low luminosity AGNs, but
their fraction increases towards intermediate and high luminosity. However,
$L_{{\rm [O \ III]}}$ of the very luminous sources in our sample is two orders
of magnitude larger than observed in the most luminous BG92 objects.
This has important implications for the physics and the structure of the NLR
in these extreme cases as discussed in the following sections.

\section{Luminosity, Equivalent-Width, and Size Correlations}
\label{LEWR}
\subsection{Measured and predicted NLR sizes}
\label{NLRsize}

B02 measured NLR sizes in seven PG quasars and compared them with sizes
obtained by Falcke et al. (1998) in nearby Seyfert 2 galaxies.
Their main finding is a strong correlation between the NLR radius
(\RNLR) and the \oiii\ line luminosity. Their relationship (B02 Eq.~1)
scaled to the somewhat different cosmology adopted here, can be written as
\begin{equation}
   R_{\rm NLR} = 2.1 L_{{\rm [O \ III]},42}^{0.52 \pm 0.06} \,\, {\rm kpc} \, ,
\label{RNLRO3}
\end{equation}
where $ L_{{\rm [O \ III]},42} = L_{{\rm [O \ III]}}/10^{42}~{\rm erg~s^{-1}}$.
This is in perfect agreement with \RNLR $\propto L_{{\rm [O \ III]}}^{1/2}$
(the uncertainty on the constant 2.1 kpc is of order 15\%).
For reasons that shall become apparent later, we prefer to use the equivalent
relation involving the \hb\ luminosity (B02 Eq.~3 converted to our
assumed cosmology)
\begin{equation}
    R_{\rm NLR} = 1.15 L_{{\rm H} \beta,42}^{0.49 \pm 0.06} \,\, {\rm kpc} \, ,
\label{RNLRHb}
\end{equation}
where $ L_{{\rm H} \beta,42} = L_{{\rm H} \beta}/10^{42}~{\rm erg~s^{-1}}$.

The recent, more detailed work of Schmitt et al. (2003) use a sample of
60 Seyfert 1 and Seyfert 2 galaxies and discuss in detail the differences
between the two sub-groups, the concentration of the \oiii\ emission, etc.
The main finding which is relevant to our work is the following
relationship (adjusted to the cosmology used here)
\begin{equation}
R_{\rm maj} \simeq 1.2 L_{{\rm [O \ III]},42}^{0.33 \pm 0.04} \,\, {\rm kpc}
\, ,
\label{schmitt}
\end{equation}
where $R_{\rm maj}$ is the size of the semi-major axis of the \oiii\
nebulosity. This is significantly different from the B02 results in both the
$R-L$ dependence and the normalization. Schmitt et al. (2003) have also
investigated the correlation when the seven B02 sources are added to their
sample. The result is a steeper dependence of the form
$R_{\rm maj} \propto L_{{\rm [O \ III]}}^{0.42}$.
A new work by Bennert et al. (2004) argues that much of the difference is due
to orientation since most of the sources in Schmitt et al. (2003) are Seyfert
2s while the more luminous sources in B02 are all type-I AGNs. We shall
return to this issue in \S~\ref{discussion}.

\begin{figure}
\epsscale{1.2}
\plotone{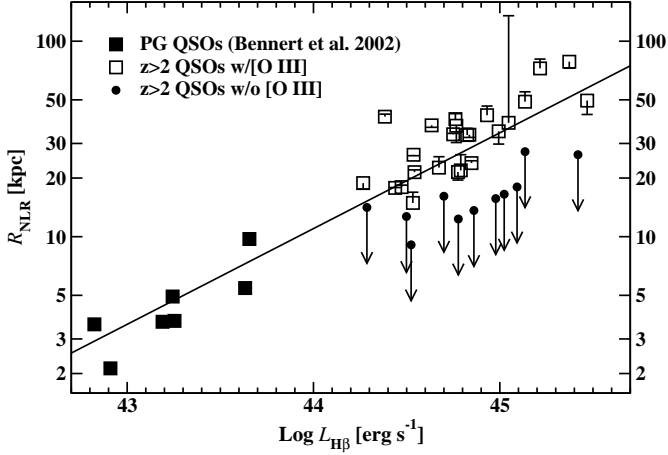}
\caption
{The $R_{\rm NLR}$ vs. $L_{{\rm H}\beta}$ diagram including our luminous
high-$z$ quasars. {\it Filled squares} are the original B02 quasar data,
and the {\it straight line} is the B02 $R_{\rm NLR}-L_{{\rm H}\beta}$ best-fit
line. {\it Empty symbols} represent high-$z$ quasars with measured
[O {\sc iii}] lines, for which $R_{\rm NLR}$ was {\em inferred} from the B02
$R_{\rm NLR}-L_{\rm [O \ III]}$ relation. One-sided error bars on the {\it
empty symbols} indicate the difference between best-fit and 'direct'
measurements (see text). {\it Arrows} represent high-$z$ quasars, for which we
have an upper limit on the \oiii\ line flux and hence
a limit on $R_{\rm NLR}$ from the B02 relation. Note the two distinct groups
and the enormous predicted \RNLR\ at high \hb\ luminosity.}
\label{RL}
\end{figure}

Theoretical suggestions that the NLR size should scale with
$L_{\rm ion}^{1/2}$, where $L_{\rm ion}$ is the ionizing source luminosity,
have been discussed in many
papers (see Netzer 1990 for references prior to 1990 and Dopita et al
2002 for more recent publications). This is based on the assumption that
both the broad line region (BLR) gas and the NLR gas are photoionized
by a central source whose spectral energy distribution changes only
slightly with source luminosity. Spectroscopic studies show a remarkable
similarity between the emission line spectrum of high and low luminosity AGNs.
This suggests that the ionization parameter, $U$ (defined here as the ratio of
the Lyman continuum photon density to the hydrogen number density $N_{\rm H}$),
and the typical gas density, are basically the same in all sources. Since
$U \propto L_{\rm ion}/N_{\rm H} R^2$, we find $R \propto L_{\rm ion}^{1/2}$.
An additional assumption is that most of the \oiii\ emission originates in
radiation-bounded clouds, because of the fairly uniform value of
I(\oiii)/I(narrow \hb). This allows to replace $L_{\rm ion}$ by the
luminosity of any hydrogen recombination line, e.g., \hb. We note that
there is a well established $R-L$ correlation for the BLR gas (e.g.,
Kaspi et al. 2000) where reverberation mappings show that
$R_{\rm BLR} \propto L^{0.6 \pm 0.1}$.

\begin{figure}
\plotone{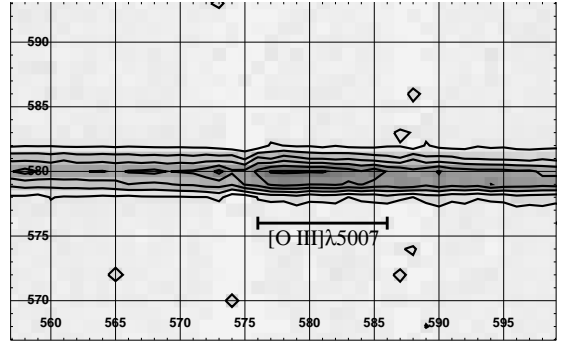}
\caption
{Two dimensional spectrum around the \oiii\ region
of 2QZ~J222006.7$-$280324. The coordinates of the vertical
and horizontal axes are given in pixels, where each pixel in the
spatial (vertical axis) corresponds to 0.446\arcsec\ ($\sim3.5$ kpc).
Note that there is no trace of extended \oiii\ emission beyond
a $\sim1$\arcsec\ ($\sim7.2$ kpc) radius from the center.}
\label{2Dspec}
\end{figure}

A real physical explanation of any $R-L$ dependence is still lacking
since the mechanism controlling the gas density and location is unknown.
Some papers assume a stratified, radiation bounded NLR with a
pre-chosen run of density and hence level of ionization (e.g., Netzer
1990; Komossa \& Schulz 1997). Such models are naturally normalized in
incident flux units ($L_{\rm ion}/R^2$) and can be tuned to produce the same
mean $U$ for all sources. However, some narrow emission lines are
probably produced in density bounded gas (e.g., Binette, Wilson, \&
Storchi-Bergmann 1996) which considerably complicate the  models.
It is also clear (e.g., Alexander et al. 1999) that well studied NLRs
contain a large range of conditions with a spread in ionization
parameter and gas density. An almost orthogonal approach is provided
by the ``locally optimally emitting cloud'' (LOC) model
(Ferguson et al. 1997). The assumption in this case  is of a large
range of conditions (density and covering factor) at each location, where
the intensities of the various lines reflect the line production
efficiency at each location. This efficiency is the highest for
densities that are close to the critical density of the line in
question. The model provides a natural scaling of $L_{\rm ion}$ with
\RNLR\ provided there is a large reservoir of gas with similar
properties in all AGNs and on all scales. Finally, there are equally
complex NLR models where shock excited gas contributes significantly
to the observed NLR emission (e.g., Contini, Prieto, \&  Viegas 1998;
Schiano 1986; Dopita and Sutherland 1995) but no natural $R-L$ scaling.

The recent papers by Dopita et al. (2002) and Groves et al. (2004)
provide a more solid foundation to NLR modeling. These authors assumed dusty,
stratified NLR clouds where the external radiation pressure acts
mostly on the dust particles and forces the local ionization parameter
to certain specific values such that in the \oiii\ producing gas,
$U\simeq10^{-2}$. The model suggests a natural \RNLR $\propto
L^{1/2}$ dependence. It also implies that the composition and
temperature of the gas are rather different from those assumed in
other models because of metal depletion.

\subsection{New {\rm [O {\sc iii}]} and \RNLR\ measurements}
\label{O3RNLR}

The combination of the B02 observations and the recent theoretical
developments point to a ``natural'' \RNLR$\propto L_{\rm ion}^{1/2}$ dependence
yet raises a severe problem regarding the NLR size in high luminosity AGNs.
Any such scaling will lead, at large enough $L_{\rm ion}$, to sizes that are
larger than galactic sizes. The B02 results addressed here are used for
normalizing this relationship, but the problem exists at high luminosity
whether or not their scaling is correct. Given this, we would not expect to
see any strong \oiii\ emitters in high luminosity sources, yet
our new observations clearly show such objects.

To define the problem in a more quantitative way we plot in Fig.~\ref{RL}
three quantities vs. $L_{{\rm H} \beta}$ (luminosity of the entire emission
line). The first is the measured \RNLR\ from the B02 sample (seven sources)
where we also show the best (modified) B02 fit (Eq.~\ref{RNLRHb}). 
The second is from our newly observed \hb\ and \oiii\ lines with two
additional sources from Dietrich et al. (2002a). For these we use
Eq.~\ref{RNLRO3} {\it to guess} \RNLR, given the observed
$L_{{\rm [O \ III]}}$.
The predicted \RNLR\ for most sources in this group lie close to the
value predicted from $L_{{\rm H} \beta}$ (the straight line) confirming the
small scatter in I(\oiii)/I(\hb). The third group includes the seven
sources from our sample and the four sources from Dietrich et
al. (2002a) where no \oiii\ has been detected. For these we use \RNLR\
derived from the upper limits on $L_{{\rm [O \ III]}}$. The upper limits on
\RNLR\ obtained in this way are a factor of $2-3$ smaller than those
derived from $L_{{\rm H} \beta}$.

The implications of Fig.~\ref{RL} are clear. For those sources with measured
\hb\ and \oiii\ lines, the derived \RNLR\ is enormous, exceeding 70 kpc in
the most luminous sources. We consider those sizes
completely unreasonable for reasons that are discussed in the next
section. Like Fig.~\ref{O3hb_hist}, this diagram suggest a dichotomy
in the properties of the high luminosity quasars, where some sources
show strong \oiii\ lines and others show no or very weak such emission.

There are two other ways to verify, experimentally, the B02 claim.
Most of our sample is at $z\simeq2.5$. At this redshift, and
the chosen cosmology, the angular diameter distance is a weak function
of redshift and corresponds to about 7 kpc per arc-sec. The
predicted \RNLR, using the B02 relationships and our measured \oiii\
luminosities, corresponds to a total extent of 2\arcsec-10\arcsec. This can
be tested by spatially resolved space and ground-based observations. The 2D
spectra of two of our sources ([HB89]~0329$-$385, and
2QZ~J222006.7$-$280324) show prominent \oiii\ emission which allow such
measurements (see Fig.~\ref{2Dspec}). In these cases, most of the line
flux ($>99$\%) is emitted within the central four pixels corresponding
to a radius of $7.2$ kpc at each source. This size is a strong upper
limit since much of the flux is likely to be due to the PSF
(corresponding to $\sim1$\arcsec\ at the time of observations). The
two upper limits on \RNLR\ obtained in this way are a factor of
$\sim5$ smaller than the radii derived from the B02 relationships.

The term ``NLR radius'' used by B02 is ambiguous since those
authors used very low surface brightness features to define the dimension of
the \oiii\ nebulosity. We used the {\sl HST} archive to extract and re-analyze
the B02 images. In particular, we examined the source showing the largest
\oiii\ nebulosity, PG~1049$-$005, and remeasured its observed \oiii\ image.
We found that 95\% of the line emission is encircled within a radius of
1.1\arcsec\ which corresponds to a radius of 5.5 kpc at the source. This  
is half of the radius deduced by B02 and suggests that the bulk of the
NLR emission is emitted within a volume which is much smaller than
inferred by their relationships. The different way of measuring the \oiii\
nebulosity is probably the main source of discrepancy in normalization
(i.e., the NLR radius at $L_{{\rm [O \ III]},42}=1$) between the 
B02 and the Schmitt et al. (2003) works.

\begin{figure}
\epsscale{1.2}
\plotone{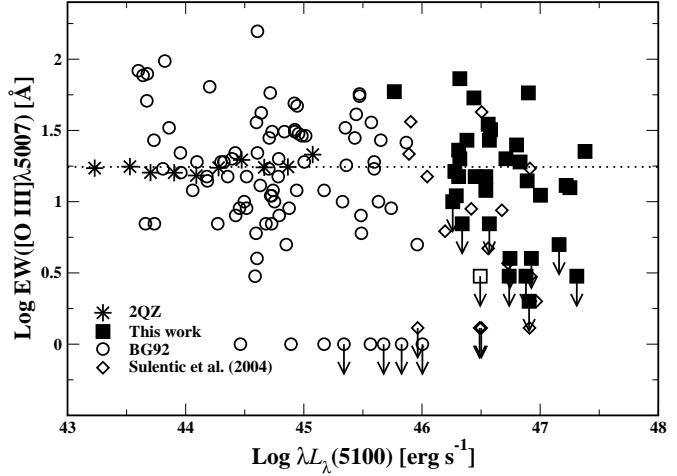}
\caption
{Baldwin relationship for \oiii. Mean (over luminosity
bins) values for the Croom et al. (2002) 2QZ sources
are marked with {\it asterisks}. The data presented in this paper
are marked with {\it filled squares}. The BG92, and Sulentic et al.
(2004) samples are marked with ({\it circles}), and ({\it diamonds}),
respectively. Upper limits on EW(\oiii) are marked with {\it arrows}).
The mean EW(\oiii) of the 2QZ sample is marked by a {\it dotted line}.}
\label{O3baldwin}
\end{figure}

In summary, our new observations contradict the B02 results in two
ways. First, about one third of the high luminosity sources show no
trace of an NLR. Second, there are direct indications in three cases,
and sound theoretical reasons (see below), to suggest 
that most of the NLR emission is restricted to a volume
which is much smaller than inferred from the B02 relationships.

\begin{figure*}
\plotone{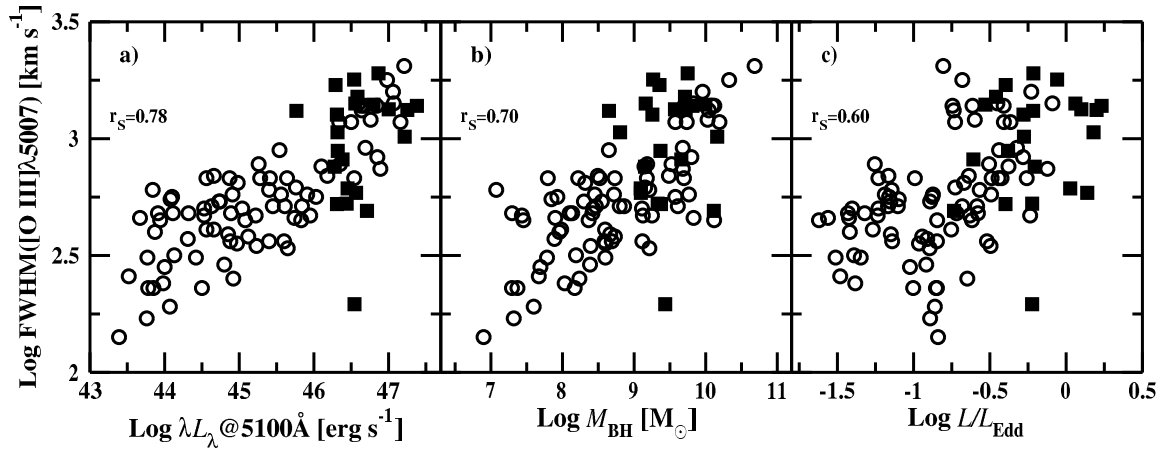}
\caption
{FWHM(\oiii) vs. {\it (a)} $\lambda L_{\lambda}(5100)$,
{\it (b)} $M_{\rm BH}$, and {\it (c)} $L/L_{\rm Edd}$ for the
Shields et al. (2003) sample ({\it empty symbols}) and the high-$z$ quasars
presented in this paper ({\it filled symbols}). The strongest correlation is
between FWHM(\oiii) and luminosity. The correlation weakens as the dependence
on luminosity drops from $L$ through $M_{\rm BH}$ ($\propto L^{0.6}$) to
$L/L_{\rm Edd}$ ($\propto L^{0.4}$), as indicated by the
Spearman rank correlation coefficients at the top left of each panel.}
\label{FWO3_L}
\end{figure*}

\subsection{The Baldwin relationship for {\rm [O {\sc iii}]}}
\label{baldwinO3}

The new data show the presence of two groups of luminous AGNs, those
with strong \oiii\ and those with no (or very weak) such line. Out of
the 35 sources investigated by us (6 from Dietrich et al. 2002a and 29
from our IR sample), 24 belong to the first group and 11 to the second.
As mentioned above, earlier studies, like the Yuan \& Wills (2003) work,
already found very weak \oiii\ in several AGNs. Most or all of these are known
BALQSOs while only two of the 10 sources discussed here (one from Dietrich et
al. 2002a, [HB89]~0105$-$265 and one, [HB89]~1246$-$057, from our sample) show
BALQSO properties. Thus, weak or no \oiii\ seems to be a common property
of many luminous AGNs. 

To further illustrate this point we plot in Fig.~\ref{O3baldwin}
EW(\oiii) versus $\lambda L_{\lambda}$(5100\AA) for the high-$z$
quasars from our sample and that of Sulentic et al. (2004), for which
$B$ magnitudes were transformed to $\lambda L_{\lambda}$(5100) assuming
$L_{\nu}\propto \nu^{-\alpha}$ with $\alpha=0.5$. We also included the BG92
sample in the diagram,
and the 2QZ data
of Croom et al. (2002), for which $b_{\rm J}$ magnitudes were transformed to
$\lambda L_{\lambda}$(5100) using the same methods as above. The diagram shows
that for the population of strong \oiii\ emitters, there is no reduction of
EW(\oiii) with source luminosity.
On the other hand, there are many weak, or no \oiii\ emitters at high
luminosity that could give the impression that the line EW decreases
with increasing source luminosity. (In fact, the Croom et al. 2002 data
represent mean EW(\oiii) in several luminosity bins. Assuming a fraction of
weak \oiii\ emitters in those bins similar to the one found by
BG92, we find that the plotted average may underestimate the
typical EW(\oiii) in those sources by about 25\%).
We suspect that the Dietrich et al. (2002b) claim of a Baldwin relationship
for this line is the result of their averaging together four very weak
\oiii\ emitters with two sources showing ``typical'' EW(\oiii).

For completeness, we tested also the correlation of EW(\oiii) with
accretion rate (in terms of the $L_{\rm Bol}/L_{\rm Edd}$
ratio, hereafter $L/L_{\rm Edd}$; see Paper I for more details)
Correlations with accretion rate are found to be extremely
important in Paper I and in \S~\ref{HbFeII} below. For EW(\oiii), this
correlation is not significant (see Table~\ref{corrmat}).

\subsection{FWHM correlations}
\label{FWO3corr}

We have tested our sample for possible correlations of luminosity, black hole
(BH) mass, and accretion-rate with FWHM(\oiii). Such correlations have been
investigated in the past (e.g., Shields et al. 2003), and FWHM(\oiii) has been
suggested as a potentially useful surrogate for $\sigma_*$.

Fig.~\ref{FWO3_L} shows FWHM(\oiii) vs. source luminosity, $M_{\rm BH}$, and
$L/L_{\rm Edd}$ for our sample and the Shields et al. (2003) sources.
To obtain the intrinsic line width we assumed
\begin{equation}
\Delta \lambda_{\rm true}^2 = \Delta \lambda_{\rm obs}^2 - \Delta
\lambda_{\rm inst}^2 ,
\label{eq:width}
\end{equation}
where $\Delta \lambda_{\rm inst}$ is the instrumental resolution. Since the
slit-width and the seeing disk sizes were comparable during the time
of observations, this is a reasonable assumption for line profiles
that are indistinguishable from a Gaussian.
The uncertainty on FWHM(\oiii) is quite large for the poorer S/N spectra and
is of the order of the instrumental (or rebinned instrumental) resolution
($\sim600$ \kms). BH masses and $L/L_{\rm Edd}$ for all sources on these
diagrams were calculated as prescribed in Paper I.

On their own, the measured FWHM(\oiii) for our high-$z$ sources show no
correlation with luminosity, BH mass, or accretion-rate, due to the narrow
luminosity range of our sample. The additional Shields et al. (2003) data
increase this range considerably and show significant correlations with all
three parameters. The strongest correlation is between FWHM(\oiii) and
luminosity. The strength of the correlation increases with the dependence on
luminosity, i.e. less significant correlations as one goes from $L$ to 
$M_{\rm BH}$ ($ \propto L^{0.6}$) to  $L/L_{\rm Edd}
(\propto L^{0.4}$).

\subsection{Correlations involving {\rm H}$\beta$ and {\rm Fe {\sc ii}}}
\label{HbFeII}

\begin{figure}
\epsscale{1.2}
\plotone{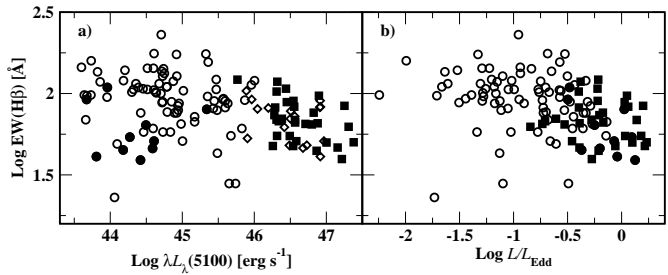}
\caption
{EW(\hb) vs. ({\it a}) luminosity and ({\it b}) $L/L_{\rm Edd}$ for the BG92
sample (including broad-line AGNs, {\it empty circles}, and NLS1s {\it filled
circles}), the Sulentic et al. (2004) sample ({\it empty diamonds} in panel
{\it a}) and the high-$z$ quasars presented in this paper
({\it filled squares}).}
\label{EW_Hb}
\end{figure}

\begin{figure}
\epsscale{1.2}
\plotone{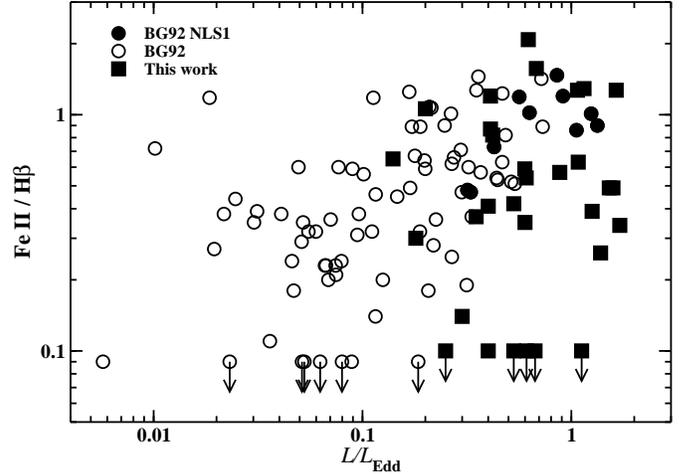}
\caption
{Fe {\sc ii}/H$\beta$ vs. $L/L_{\rm Edd}$ for the BG92 sample,
and the high-$z$ quasars presented in this paper (symbols are
as in Fig.~\ref{EW_Hb}).
{\it Arrows} represent no detection of iron emission (in the BG92 sample the
upper limit on the Fe {\sc ii}/\hb\ ratio was set to
0.09, which is the lowest value they reported, and in our sample the upper
limits are given in Table~\ref{fit}).}
\label{E_R_Fe2}
\end{figure}

In Paper I and in Table~\ref{fit} we give $L_{{\rm H} \beta}$, which we
transformed to
EW(\hb), and \ion{Fe}{2}/\hb. Here, again, we have tested the correlations of
these quantities against $L$, $M_{\rm BH}$, and $L/L_{\rm Edd}$ using our
sample, and the samples of Sulentic et al. (2004) and BG92 (accretion rates for
the BG92 sample were calculated as prescribed in Paper I).
For EW(\hb) we find significant correlations with both $L$ and $L/L_{\rm Edd}$.
Details of the correlations are given in Table~\ref{corrmat} and in
Fig.~\ref{EW_Hb}. In the case of \ion{Fe}{2}/\hb\ the only significant
correlation is with $L/L_{\rm Edd}$. Details of those
correlations are also given in Table~\ref{corrmat}.
Fig.~\ref{E_R_Fe2} shows the strongest correlation of \ion{Fe}{2}/\hb\
(against the accretion rate) for our new sources combined with those of BG92.

\begin{deluxetable}{lccccc}
\tablecolumns{6}
\tablewidth{0pt}
\tablecaption{Spearman Rank Correlation Coefficients Matrix
\label{corrmat}}

\tablehead
{
\colhead{Property vs.} &
\colhead{$M_{\rm BH}$} &
\colhead{$L/L_{\rm Edd}$} &
\colhead{Fe {\sc ii}/\hb\ } &
\colhead{EW([O {\sc iii}])} &
\colhead{EW(\hb)} \\
\colhead{(1)} &
\colhead{(2)} &
\colhead{(3)} &
\colhead{(4)} &
\colhead{(5)} &
\colhead{(6)}
}
\startdata
$\lambda L_{\lambda}$(5100)&{\bf 0.85}\tablenotemark{a}
&{\bf 0.48}\tablenotemark{a}& -0.04\tablenotemark{b}     &
-0.09\tablenotemark{d}&{\bf -0.37}\tablenotemark{g} \\
$M_{\rm BH}$          &\nodata & 0.03\tablenotemark{a}
&-0.22\tablenotemark{c}&
0.14\tablenotemark{e}& -0.15\tablenotemark{h} \\
$L/L_{\rm Edd}$       &\nodata &\nodata &
{\bf 0.48}\tablenotemark{c}&
-0.13\tablenotemark{e}&{\bf -0.47}\tablenotemark{h} \\
Fe {\sc ii}/\hb\      &\nodata &\nodata &\nodata    &
{\bf -0.39}\tablenotemark{f}& {\bf -0.33}\tablenotemark{b} \\
EW([O {\sc iii}])             &\nodata &\nodata &\nodata &
\nodata    &{\bf 0.28}\tablenotemark{d} \\
EW(\hb)               &\nodata &\nodata &\nodata &\nodata &\nodata \\
\enddata

\tablecomments{Significant correlations with chance probabilities smaller than
1\% are given in {\em bold face}. Upper limits were not included in the
correlations.}
\tablenotetext{a}{118 sources from BG92 and this work}
\tablenotetext{b}{124 sources from BG92, Sulentic et al. (2004), and this work}
\tablenotetext{c}{107 sources from BG92 and this work}
\tablenotetext{d}{121 sources from BG92, Sulentic et al. (2004), and this work}
\tablenotetext{e}{104 sources from BG92 and this work}
\tablenotetext{f}{110 sources from BG92, Sulentic et al. (2004), and this work}
\tablenotetext{g}{135 sources from BG92, Sulentic et al. (2004), and this work}
\tablenotetext{h}{118 sources from BG92 and this work}

\end{deluxetable}

The Baldwin relationship found here for EW(\hb), and shown in Fig.~\ref{EW_Hb}
is interesting since it is in contradiction with the Croom et al. (2002)
finding for this line. However, our
combined sample is far from being complete, in particular at the high
luminosity range where we specifically chose to observe high luminosity
sources. Thus it is likely that lower luminosity sources at high redshift
would have a larger EW(\hb) that will spoil the correlation. Regarding the
dependence on accretion rate, similar selection effects may be operating but
the observed correlation is so strong that we suggest that this may be a real
effect. We also note that in Fig.~\ref{EW_Hb}, the narrow-line Seyfert 1
galaxies (NLS1s) are situated very
close to the high luminosity, high accretion rate sources. The situation
resembles the Baskin \& Laor (2004) finding that EW(\CIV) depends strongly on
$L/L_{\rm Edd}$ even in low luminosity AGNs that do not show the Baldwin
relationship for this line.

Fig.~\ref{E_R_Fe2} is hard to interpret. Considering only sources with
measurable \ion{Fe}{2}/\hb\ (i.e., a ratio larger than $\sim0.1$) we find
a clear trend of increased \ion{Fe}{2}/\hb\ for higher accretion rates
(see Table~\ref{corrmat}).
In particular, we note the similar location on the diagram of our high-$z$
quasars and the NLS1s. However, there is a significant number of sources
with very small \ion{Fe}{2}/\hb\ and very large $L/L_{\rm Edd}$. We note,
in this respect, that our new observations extend the measurement of
\ion{Fe}{2}/\hb\ to values of $L/L_{\rm Edd}$ never investigated before.
The only exceptions, perhaps, are a few BALQSOs in the Yuan \& Wills (2003)
sample with spectral properties very different from those of our (non-BAL)
sources (note that these authors measured the entire optical \ion{Fe}{2} blends
and their values must be scaled down by a factor 3.58 for comparison with the
measurements presented in our work; B. Wills private communication).

The above two findings suggest that the main reason for the increase of
\ion{Fe}{2}/\hb\ with the accretion rate is the decrease of EW(\hb) with
$L/L_{\rm Edd}$. In fact, looking at EW(\ion{Fe}{2}) against $L/L_{\rm Edd}$
(not shown here)
we find no correlation at all. The fractional increase in \ion{Fe}{2}/\hb\
with accretion rate is also consistent with the decrease in EW(\hb).
However, we cannot rule out the possibility that changes in the Fe/H abundance
ratio are involved too, as shown, in Paper I, to be the case for N/C.

\begin{figure}
\epsscale{1.2}
\plotone{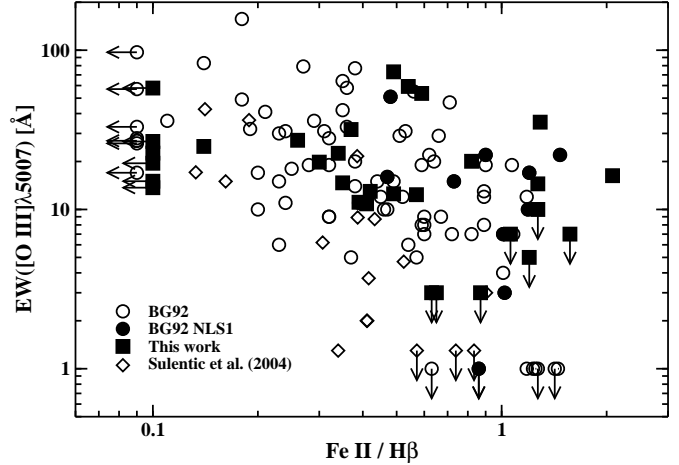}
\caption
{EW(\oiii) vs. Fe {\sc ii}/H$\beta$ for the BG92 sample, the Sulentic
et al. (2004) sample, and the high-$z$ quasars presented in this paper
(symbols are as in Fig.~\ref{EW_Hb}).
Upper limits on Fe {\sc ii}/\hb\ ({\it arrows pointing left}) are similar
to those shown in Fig.~\ref{E_R_Fe2}. Upper limits on EW(\oiii), i.e., no
\oiii\ detection, for the BG92 (Sulentic et al. 2004) sample were set to 1
(1.3) \AA, respectively, which is the lowest value each of them reported. The
upper limits on EW(\oiii) for our sample are given in Table~\ref{fit}.}
\label{R_Fe2_W_O3}
\end{figure}

In Fig.~\ref{R_Fe2_W_O3} we plot the well known anti-correlation between
EW(\oiii) and \ion{Fe}{2}/\hb\ (e.g., BG92) for our sources, for the low-$z$
BG92 sources, and for the new intermediate-$z$ sample of Sulentic et al.
(2004). Our sources are consistent with this trend. \ion{Fe}{2}/\hb\ for our
sources is generally higher than in BG92 and Sulentic et al. (2004)
indicating, perhaps, the strong dependence of this property on
the total luminosity. However, there are clear exceptions, i.e., sources
with very high luminosity, yet relatively small \ion{Fe}{2}/\hb. Here, again,
we note similar large values of \ion{Fe}{2}/\hb\ in several
of the BALQSOs of Yuan \& Wills (2003) .

\section{Discussion}
\label{discussion}
\subsection{Do enormous NLRs really exist?}
\label{enormous}

A major goal of the present investigation is to test the NLR properties and the
NLR spectrum in very high luminosity sources where the theoretically predicted
\RNLR, as well as the empirical B02 relationship, results in unreasonably large
dimensions. The B02 sample already included claims for sources with
\RNLR$\sim10$ kpc and our bright \oiii\ emitters would continue this
relationship to enormously large NLRs (more than 100 kpc in diameter,
see Fig.~\ref{RL}). There are, however, fundamental problems in this suggested
size-luminosity interpretation on both observational and theoretical grounds.

B02 suggested two estimates of the NLR size
(Eq.~\ref{RNLRO3}~\& ~\ref{RNLRHb}).
We separate the discussion of those claims into two, according to the two
sub-groups discovered here (the weak and the strong \oiii\ emitters).
For the weak \oiii\ emitters we find a clear contradiction between the
\RNLR\ based on the observed \hb\ luminosity and the one based on the observed
(or the upper limit on) \oiii\ luminosity (Fig.~\ref{RL}). As for the strong
\oiii\ emitters, 2D spectra of two of our sources rule out the large predicted
dimensions (\S~\ref{O3RNLR}). Moreover, our new measurements of the largest
\oiii\ nebula in the B02 sample are also in conflict with the B02
relationships. Based on the evidence in hand we suspect that the \RNLR$\propto
L_{\rm ion}^{1/2}$ dependence breaks down at some intermediate luminosity scale
and that the ``true'' NLR radius, defined here as the radius encompassing
95\% of the line emission, does not exceed a few kpc even in the most luminous
quasars.

There are other predictions that make us question the existence of such
enormous NLRs. The suggested \RNLR$\propto L_{\rm ion}^{1/2}$ relationship
would predict NLR sizes, in the most luminous AGNs, that exceed the size of
the largest known bulges and, in fact, the size of the largest known
galaxies (except for some cD galaxies). Such sizes are unacceptable
for several reasons. First, the escape velocity from a spherical
galaxy is roughly $290 M_{11}^{1/2} R_{10\,{\rm kpc}}^{-1/2}$ \kms, 
where $M_{11}$ is the mass in units of $10^{11}$ \Msun\ and $R_{10}$
the radial distance in units of 10 kpc. The new FWHMs listed in Table~\ref{fit}
and shown in Fig.~\ref{FWO3_L}, compared with the predicted \RNLR, suggest
therefore, dynamically unbounded NLRs.
Given those velocities and dimensions, the dynamical time for the most
luminous \oiii\ emitters, is a few $\times 10^7 R_{10}$ years
suggesting a short lived phenomenon. The inferred amount of ionized
gas in such NLRs, given radiation bounded gas and a ``typical'' ionization
parameter, is
\begin{equation}
M_{\rm NLR} \simeq 10^9 \left [ \frac{C_f}{0.1} \right ] R_{10}^2 N_{21}
\,\, M_{\odot}  \, ,
\label{eq:MNLR}
\end{equation}
where $C_f$ is the covering fraction and $N_{21}$ the column density
in units of $10^{21}$ \cmii. Simple photoionization arguments show
that for strong \oiii\ emitters $N_{21}>1$. This would give 
$M_{\rm NLR} > 10^9$ \Msun\ for all of our sources and $M_{\rm NLR} > 10^{10}$
\Msun\ for the most luminous \oiii\ emitters. 
Judging by the observed FWHM, most of this material is probably
unbound and thus flows from the center at outflow rates approaching
$10^6 \ R_{10}^2 \ \Msun\ {\rm yr}^{-1}$ assuming a density of
$\sim10^3 \ {\rm cm}^{-3}$. All those numbers seem
incompatible with long term mass ejection in AGN-hosting galaxies. 

The Schmitt et al. (2003) results alleviate some of these difficulties
because of the smaller size normalization and the flatter
\RNLR-$L_{{\rm [O \ III]}}$ dependence. However, it is not at all clear that
extrapolating their results to much higher luminosity with the suggested
$L^{1/3}$ dependence (Eq.~\ref{schmitt}) is justified in view of the B02
measurements.

We are facing a situation where sound physical arguments (the narrow line
spectrum, the Groves et al. 2004 dusty NLR model, etc.) support the expected
\RNLR$\propto L^{1/2}$ relationship, yet its application to the most luminous
quasars give unreasonably large sizes, masses and mass outflow rates.
As explained below, the most likely explanation in our opinion is that
``typical'' NLRs (i.e., those similar in their properties to the ones observed
in Seyfert 1 galaxies) cannot last very long in high and perhaps also in
intermediate luminosity AGNs. This means that the \oiii\ emitting regions 
in our sample may be of a different origin and physical properties.

\subsection{Luminous NLRs as star forming regions}
\label{starburst}

If compact NLRs are indeed typical of many high luminosity AGNs, then
their properties must be very different from the properties of those NLRs
observed in nearby sources. In particular, the gas density
in the highest luminosity NLRs must be several orders of magnitude larger.
Consider for example a maximum NLR size of $\sim3$ kpc and assume a similar
ionization parameter in all NLRs. This would mean that the gas density in the
most luminous \oiii\ emitters is $10^2-10^3$ times larger than the
density in nearby Seyfert 1 galaxies. The spectroscopic properties 
must be very different too which can, in principle, be tested by
accurate observations. Unfortunately, present day IR spectroscopy is
very limited in this respect because of the restricted wavelength 
bands available to ground-based observations. 

A possible origin of a high density gas in kpc-scale nuclear regions
is violent star-forming activity. Such events can produce
high density, large column density, non-solar composition dusty
gas. The overall spectrum of such regions is likely to differ from the
spectrum of nearby, lower density NLRs. At present we can only observe
\oiii\  which is the strongest emission line under a variety of
conditions. Future, space-born spectroscopy, will be able to test this idea
by looking for other emission lines.

The scenario we propose to explain the observations of our high
luminosity AGNs, and the apparent break down of the
\RNLR$\propto L_{\rm ion}^{1/2}$ relationship, is of two distinct populations.
One where such scaling continues to high luminosity due to radiation pressure
force or other effects. This results in short-lived enormous NLRs that will
show basically no nuclear narrow emission lines during most of their life.
The other group is those sources where
starburst or another unknown process ejects high density gas into their nuclear
region. This gas is ionized and excited by the central radiation source and
produces the observed strong \oiii\ lines. Such ``star-forming NLRs''
would have spectral properties that are rather different from those
observed in nearby less luminous sources. Given the similar fraction of weak
\oiii\ emitters in BG92 and in our new sample, we suggest
that the phenomenon is of a continuous nature and starts at some intermediate
luminosity. Its clearest manifestation is in the highest luminosity sources
such as the ones observed here.

\subsection{Correlations involving {\rm Fe {\sc ii}}, {\rm H}$\beta$, and
{\rm [O {\sc iii}]}}
\label{FeIIHbO3corr}

The sample used here, which combines data from various different sources, is
not complete. However, it allows the best test, so far, of the
BG92 \ion{Fe}{2}/\hb\ --\oiii\ relationship and several other suggested
correlations at the high end of the AGN luminosity function.
The results presented here suggest that the most extreme
values of \ion{Fe}{2}/\hb\ require very high luminosity. They also suggest that
higher accretion rate results in larger \ion{Fe}{2}/\hb\ (Fig.~\ref{E_R_Fe2})
and that $L/L_{\rm Edd}$ is the most important factor for determining several
other correlations. There are, however, some exceptions, i.e., sources
with large $L/L_{\rm Edd}$ yet small \ion{Fe}{2}/\hb. 
Our work shows that the decrease of EW(\hb) with increasing $L/L_{\rm Edd}$
in the present sample is
probably the cause for the increase of \ion{Fe}{2}/\hb\ with the accretion
rate. EW(\ion{Fe}{2}) by itself does not depend on the accretion rate and it
is therefore possible that the iron-to-hydrogen abundance ratio plays no role
in this correlation. On the other hand, we do not have the data or a good
enough theory (e.g., Verner et al. 2004 and references therein) to completely
rule out the possibility that the iron abundance depends on luminosity or on
the accretion rate. In a similar way, EW(\oiii) is {\em not} correlated with
$L/L_{\rm Edd}$, but we suggest that the \ion{Fe}{2}/\hb\ -- EW(\oiii)
anti-correlation is driven by the accretion rate.

Sulentic et al. (2004) and others argued that the general AGN population should
be divided into two sub-groups according to the so-called Eigenvector 1
(E1, see BG92) which contains four main observables. According to this scheme,
group A sources show the strongest E1 properties. Most radio-loud AGNs belong
to group B, i.e., those with weaker E1 properties.
Here we are not interested in E1 as such since our sample is not complete and
so is the combination of our sample with the BG92 and the Sulentic et al.
(2004) samples. We are however in a position to look at the extreme end of the 
distribution in $L$ and in $L/L_{\rm Edd}$ and test for those quantities
that depend on them.
We have already discussed all the relevant correlations. Here we note that
all those involving $L/L_{\rm Edd}$ show that NLS1s occupy the same part of
parameter space as the high luminosity, high accretion rate AGNs observed by
us. These sources are thus the high luminosity analogs of NLS1s and should,
perhaps, be referred to as narrow-line type 1 quasars (NLQ1s).

\section{Conclusions}
\label{conclusions}

We have discussed the near-IR spectra of 29 newly observed high-luminosity
high-redshift AGNs and used the data to argue that previous claims for
expected and observed \RNLR$\propto L_{{\rm [O \ III]}}^{1/2}$ dependence are
in conflict with the observations. About 2/3 of all very high luminosity
sources show strong \oiii\ lines while the remaining objects show no or very
weak such line. We suggest that the NLR properties in high luminosity, strong
\oiii\ emitters are very different from those observed in nearby AGNs, and
possibly imply denser than ``usual'' NLRs.
The origin of the high density gas is likely to be a violent star-forming event
in the nucleus. We also investigated \ion{Fe}{2}/\hb\ and EW(\oiii) at the
high end of the luminosity and $L/L_{\rm Edd}$ in  AGNs and showed that the
first of those is probably driven by the overall accretion rate while the
second is independent of source luminosity (i.e., no Baldwin relationship) or
accretion rate. The \ion{Fe}{2}/\hb\ ratio may be an iron abundance indicator
but this cannot be proven observationally because of the EW(\hb) dependence on
accretion rate.

\acknowledgements

We are grateful to the technical staff at the AAT and TNG
observatories for invaluable help during the observations.
We acknowledge constructive remarks made by an anonymous referee, which helped
to improve this work.
This work is based on observations made with the Italian Telescopio Nazionale
Galileo (TNG) operated on the island of La Palma by the Centro Galileo Galilei
of the INAF (Istituto Nazionale di Astrofisica) at the Spanish Observatorio 
del Roque de los Muchachos of the Instituto de Astrofisica de Canarias.
We would like to thank Angela Cotera for allowing us to use half a night of
her time at AAT, and Dirk Grupe for providing us an electronic version of
the Boroson \& Green (1992) Fe {\sc ii} template.
We gratefully acknowledge constructive remarks from an anonymous referee,
who helped to improve this work considerably.
The 2dF QSO Redshift Survey (2QZ) was compiled by the 2QZ survey team
from observations made with the 2-degree Field on the Anglo-Australian
Telescope.
Funding for the creation and distribution of the SDSS Archive has been
provided by the Alfred P. Sloan Foundation, the Participating
Institutions, the National Aeronautics and Space Administration, the
National Science Foundation, the U.S. Department of Energy, the
Japanese Monbukagakusho, and the Max Planck Society.
The SDSS Web site is http://www.sdss.org/.
The SDSS is managed by the Astrophysical Research Consortium (ARC) for
the Participating Institutions. The Participating Institutions are The
University of Chicago, Fermilab, the Institute for Advanced Study, the
Japan Participation Group, The Johns Hopkins University, Los Alamos
National Laboratory, the Max-Planck-Institute for Astronomy (MPIA),
the Max-Planck-Institute for Astrophysics (MPA), New Mexico State
University, University of Pittsburgh, Princeton University, the United
States Naval Observatory, and the University of Washington.
This research has made use of the NED database which is operated by
the Jet Propulsion Laboratory, California Institute of Technology,
under contract with the National Aeronautics and Space Administration.
This work is supported by the Israel Science Foundation grant 232/03.
RM acknowledges partial support by the Italian Ministry of
Research (MIUR).


\begin{thebibliography}{}

\bibitem[Alexander et al.(1999)]{1999ApJ...512..204A} Alexander, T., Sturm, 
E., Lutz, D., Sternberg, A., Netzer, H., \& Genzel, R.\ 1999, \apj, 512, 204
\bibitem[Baldwin(1977)]{1977ApJ...214..679B} Baldwin, J.~A.\ 1977, \apj, 
214, 679 
\bibitem[Barth et al.(2001)]{2001ApJ...546..205B} Barth, A.~J., Ho, L.~C., 
Filippenko, A.~V., Rix, H., \& Sargent, W.~L.~W.\ 2001, \apj, 546, 205
\bibitem[Baskin \& Laor(2004)]{2004MNRAS.350L..31B} Baskin, A.~\& Laor, A.\ 
2004, \mnras, 350, L31
\bibitem[Bennert et al.(2002)]{2002ApJ...574L.105B} Bennert, N., Falcke, 
H., Schulz, H., Wilson, A.~S., \& Wills, B.~J.\ 2002, \apjl, 574, L105
\bibitem[Bennert et al.(2004)]{2004} Bennert, N., Falcke, 
H., Shchekinov, Y., \& Wilson, A.~S. 2004, in
The Interplay among Black Holes, Stars and ISM in Galactic Nuclei, IAU 222,
eds. T. Storchi-Bergmann, L.C. Ho, and H.R. Schmitt
\bibitem[Binette, Wilson, \& Storchi-Bergmann(1996)]{1996A&A...312..365B} 
Binette, L., Wilson, A.~S., \& Storchi-Bergmann, T.\ 1996, \aap, 312, 365
\bibitem[Boroson \& Green(1992)]{1992ApJS...80..109B} Boroson, T.~A.~\& 
Green, R.~F.\ 1992, \apjs, 80, 109
\bibitem[Contini, Prieto, \& Viegas(1998)]{1998ApJ...492..511C} Contini, 
M., Prieto, M.~A., \& Viegas, S.~M.\ 1998, \apj, 492, 511 
\bibitem[Croom et al.(2002)]{2002MNRAS.337..275C} Croom, S.~M.~et al.\ 
2002, \mnras, 337, 275
\bibitem[Dietrich, Appenzeller, Vestergaard, \& Wagner(2002)]
{2002ApJ...564..581D} Dietrich, M., Appenzeller, I., 
Vestergaard, M., \& Wagner, S.~J.\ 2002a, \apj, 564, 581
\bibitem[Dietrich et al.(2002)]{2002ApJ...581..912D} Dietrich, M., Hamann, 
F., Shields, J.~C., Constantin, A., Vestergaard, M., Chaffee, F., Foltz, 
C.~B., \& Junkkarinen, V.~T.\ 2002b, \apj, 581, 912
\bibitem[Dopita \& Sutherland(1995)]{1995ApJ...455..468D} Dopita, M.~A.~\& 
Sutherland, R.~S.\ 1995, \apj, 455, 468
\bibitem[Dopita et al.(2002)]{2002ApJ...572..753D} Dopita, M.~A., Groves, 
B.~A., Sutherland, R.~S., Binette, L., \& Cecil, G.\ 2002, \apj, 572, 753
\bibitem[Falcke, Wilson, \& Simpson(1998)]{1998ApJ...502..199F} Falcke, H., 
Wilson, A.~S., \& Simpson, C.\ 1998, \apj, 502, 199
\bibitem[Ferguson, Korista, Baldwin, \& Ferland(1997)]{1997ApJ...487..122F} 
Ferguson, J.~W., Korista, K.~T., Baldwin, J.~A., \& Ferland, G.~J.\ 1997, 
\apj, 487, 122 
\bibitem[Groves]{2004Apj} Groves, B.~A., Dopita, M.~A., \& Sutherland, R.~S.
2004 (submitted to ApJ)
\bibitem[Kaspi et al.(2000)]{2000ApJ...533..631K} Kaspi, S., Smith, P.~S., 
Netzer, H., Maoz, D., Jannuzi, B.~T., \& Giveon, U.\ 2000, \apj, 533, 631 
\bibitem[Komossa \& Schulz(1997)]{1997A&A...323...31K} Komossa, S.~\& 
Schulz, H.\ 1997, \aap, 323, 31
\bibitem[McIntosh et al.(1999)]{1999ApJ...514...40M} McIntosh, D.~H., 
Rieke, M.~J., Rix, H.-W., Foltz, C.~B., \& Weymann, R.~J.\ 1999, \apj, 514, 40
\bibitem[Nelson(2000)]{2000ApJ...544L..91N} Nelson, C.~H.\ 2000, \apjl, 
544, L91 
\bibitem[Nelson \& Whittle(1996)]{1996ApJ...465...96N} Nelson, C.~H.~\& 
Whittle, M.\ 1996, \apj, 465, 96
\bibitem[Netzer(1990)]{} Netzer, H. 1990, in Active Galactic Nuclei,
ed., T. J.-L. Courvoisier \& M. Mayor (Berlin: Springer), 57
\bibitem[Netzer(2003)]{2003ApJ...583L...5N} Netzer, H.\ 2003, \apjl, 583, L5 
\bibitem[Schiano(1986)]{1986ApJ...302...81S} Schiano, A.~V.~R.\ 1986, \apj, 
302, 81
\bibitem[Schmitt et al.(2003)]{2003ApJ...597..768S} Schmitt, H.~R., Donley, 
J.~L., Antonucci, R.~R.~J., Hutchings, J.~B., Kinney, A.~L., \& Pringle, 
J.~E.\ 2003, \apj, 597, 768
\bibitem[Shemmer et al.(2004)]{2004ApJ...???...???} Shemmer, O., Netzer, H.,
Maiolino, R., Croom, S., Oliva, T., \& Di Fabrizio, L. \ 2004, \apj,
in press (Paper I)
\bibitem[Shields et al.(2003)]{2003ApJ...583..124S} Shields, G.~A., 
Gebhardt, K., Salviander, S., Wills, B.~J., Xie, B., Brotherton, M.~S., 
Yuan, J., \& Dietrich, M.\ 2003, \apj, 583, 124
\bibitem[Sulentic et al.(2004)]{2004???...???..???S} Sulentic, J.~W.,
Stirpe, G.~M., Marziani, P., Zamanov, R., Calvani, M., \& Braito, V. \
2004, A\&A, in press (astro-ph/0405279)
\bibitem[Veilleux(1991)]{1991ApJS...75..383V} Veilleux, S.\ 1991, \apjs, 
75, 383
\bibitem[Veilleux \& Osterbrock(1987)]{1987ApJS...63..295V} Veilleux, S.~\& 
Osterbrock, D.~E.\ 1987, \apjs, 63, 295
\bibitem[Verner et al.(2004)]{2004ApJ???...???..???V} Verner, E.,
Bruhweiler, F., Verner, D., Johansson, S., Kallman, T., \& Gull, T.
2004, ApJ, in press
\bibitem[Yuan \& Wills(2003)]{2003ApJ...593L..11Y} Yuan, M.~J.~\& Wills, 
B.~J.\ 2003, \apjl, 593, L11

\end{thebibliography}
\end{document}